\documentclass[pre,10pt,twocolumn,showpacs]{revtex4-1}
\UseRawInputEncoding
\usepackage{amsmath}
\usepackage{latexsym}
\usepackage{amssymb}
\usepackage{lipsum}
\usepackage{mathtools}
\usepackage{braket}
\usepackage{graphicx}
\usepackage[caption=false]{subfig}
\usepackage{wrapfig}
\usepackage{dsfont}
\usepackage{etoolbox}
\usepackage[colorlinks=true, citecolor=blue, urlcolor=blue]{hyperref}
\usepackage{float}
\usepackage{amsfonts}
\usepackage{soul}

\newcommand{\be}{\begin{equation}}
\newcommand{\ee}{\end{equation}}
\newcommand{\ba}{\begin{eqnarray}}
\newcommand{\ea}{\end{eqnarray}}

\DeclarePairedDelimiter\abs{\lvert}{\rvert}

\def\<{\langle}
\def\>{\rangle}

\begin{document}
\title{Thermodynamics and the quantum speed limit in the non-Markovian regime}

\author{Arpan Das}
\email{arpand@umk.pl}
\author{Anindita Bera}
\author{Sagnik Chakraborty}
\author{Dariusz Chru{\'s}ci{\'n}ski}
\affiliation{Institute of Physics, Faculty of Physics, Astronomy and Informatics,
Nicolaus Copernicus University, Grudziadzka 5/7, 87-100 Toru{\'n}, Poland}

\begin{abstract}

Quantum speed limit (QSL) for open quantum systems in the non-Markovian regime is analyzed. We provide a lower bound for the time required to transform an initial state to a final state in terms of thermodynamic quantities such as the energy fluctuation, entropy production rate and dynamical activity. Such bound was already analyzed for Markovian evolution satisfying detailed balance condition. Here we generalize this approach to deal with arbitrary evolution governed by time-local generator. Our analysis is illustrated by three paradigmatic examples of qubit evolution:  amplitude damping, pure dephasing, and the eternally non-Markovian evolution.
\end{abstract}

\maketitle

\section{Introduction}

Quantum speed limit (QSL) which sets the lower bound on the minimum evolution time of a quantum system, has been subjected to a rigorous amount of works in recent years. Historically QSL was first reported as an interpretation of time-energy uncertainty relation for quantum systems undergoing unitary dynamics.
 Mandelstam and Tamm \cite{MT} were the first to realize that the time in the time-energy uncertainty relation actually refers to the minimal time of unitary evolution between two orthogonal states $\ket{\psi(0)}$ and $\ket{\psi(\tau)}$.
They derived the first result in this direction, which is  known as the MT-bound, i.e. $\tau\geq\tau_{\rm QSL}= \pi\hbar/2\Delta E$, where ${(\Delta E)}^2$ is the variance of the energy with respect to an initial state $\ket{\psi(0)}$, undergoing an unitary evolution generated by a time independent Hamiltonian $H$. More than fifty years later, for the same scenario above, Margolus and Levitin~\cite{ML} derived another QSL formula  $\tau_{\rm QSL}=\pi\hbar/2E$, where $E$ denotes the expectation value of the Hamiltonian $H$. Together, these two results give a tight lower bound of time \cite{LT} for unitary evolution generated by a time independent Hamiltonian between two orthogonal pure states, $\tau_{\rm QSL}=\max\{\pi\hbar/2\Delta E, \pi\hbar/2E\}$.

Soon after,  many generalizations and applications followed (cf. the review \cite{reviewQSL}). QSL now has been found for
scenarios, where the two states are no longer orthogonal, 
where the two states are mixed, the Hamiltonian is time dependent, and, even for dynamical maps arising from open quantum systems \cite{TaddeiQSL, CampoQSL, DeffnerQSL}. Interestingly, QSL is not intrinsic to quantum systems only. Recently, equivalent bounds have been explored \cite{classical-qsl2, classical-qsl1, classical-qsl3, classical-qsl4} in both closed and open classical systems.
Also in recent years, the connection between non-Markovianity \cite{rivas14quantum,b-review} and quantum speed limit has received a lot of attention \cite{DeffnerQSL, non-qsl3, non-qsl4, non-qsl5, non-qsl1,non-qsl2} (and references within).
Moreover geometrical structure of quantum states allowed for even further generalization of the notion of QSL \cite{reviewQSL, geoPRX, modi1, modi2}.
Observing that distingushability of two quantum states is effectively a distance measure, QSL was found as the geodesic time required to cover a  generalized distance between two states connected by a time evolution operator. Note here, that time evolution operator in this case can even be dynamical maps of open quantum systems.
The only issue with this approach is that in quantum scenario there are infinite choices for distance measures, which makes it difficult to assess the tightest bound.

Unlike its unitary counterpart where the QSL relates to the physical terms like variance or expectation value of energy, QSL for open quantum dynamics involve abstract geometrical quantities which hardly give any physical insight into the underlying factors limiting the speed of evolution. Recently, the authors in the Ref. \cite{funoQSL} explored the physical meaning of the terms which arise in the expression of QSL and connected them to thermodynamic quantities in a scenario where quantum systems are governed by time-local  master equation \cite{breuer02}. Specifically, in the QSL they identified three distinct terms, which are energy fluctuations, a combined term of entropy production rate and dynamical activity and a third term which is purely quantum mechanical related to the velocity of the bath induced unitary dynamics.

The main motivation of this paper is to extend this formalism to include master equations which involve non-Markovian effect. We start with the most general time-local  master equation  and follow the framework of re-normalized transition rate \cite{leggio} to study the QSL. Specifically, the Pauli rate equation which might be assigned to the time-local master equation does in general involve transition rates which are negative. One can rearrange the rate equation and define renormalized transition rates such that they are always non-negative. Following this framework one can provide a sound definition of entropy production rate which in turn can be used to derive a speed limit in terms of thermodynamic quantities. We consider three physical models of qubit evolution to exemplify our approach: amplitude damping,  pure dephasing, and eternally non-Markovian  evolution. Moreover, we analyze how tight is the derived  QSL by comparing it with the original QSL defined in terms of the geometrical objects.

The paper is organized as follows. In Sec.~\ref{sec-1}, we discuss the QSL for Markovian dynamics. Time-local master equation and the framework of renormalized transition rates are introduced  in Sec.~\ref{sec-2}. We then derive the QSL in terms of thermodynamic quantities in Sec.~\ref{sec-3}. In Sec.~\ref{sec-4}, we illustrate our results numerically for amplitude damping, pure dephasing and eternally non-Markovian model, respectively. Finally, in Sec.~\ref{sec-7} we conclude.

\section{QSL for Markovian dynamics}
\label{sec-1}
Geometrical structure of quantum states \cite{KAROL} presents an elegant way to understand the idea of QSL. As the set of density matrices for a quantum system is a Riemannian manifold, the geodesic distance $D\Big(\rho(0),\rho(\tau)\Big)$  between two states $\rho(0)$ and $\rho(\tau)$ with respect to a contractive metric $g_{\mu\nu}$ is the shortest path connecting $\rho(0)$ and $\rho(\tau)$, that is,
\begin{equation}\label{D1}
D\Big(\rho(0),\rho(\tau)\Big) \leq \int_0^\tau ds ,
\end{equation}
 where $ds = v_t dt$, and $v_t = \sqrt{\sum_{\mu,\nu}g_{\mu\nu}\dot{\lambda}^{\mu}\dot{\lambda}^{\nu}}$ is a speed of the evolution along the trajectory $t \to \rho(t)$ measured w.r.t. the Riemannian metric $g_{\mu\nu}$ (here $\lambda^\mu$ denotes coordinates on the Riemannian manifold of quantum states).  The above inequality  immediately implies,
\begin{equation}   \label{A}
\tau  \geq \frac{D\Big(\rho(0),\rho(\tau)\Big)}{\langle v_t \rangle_\tau},
\end{equation}
where, $\langle v_t \rangle_\tau=1/\tau \int_0^\tau v_t dt$ denotes the time average of the speed $v_t$ over the time interval $[0,\tau]$. One can, therefore, interpret the denominator of the above equation as an average velocity.

As depicted by the authors Morozova, \v{C}encov, Petz, Hasegawa, and Hiai in Refs.~\cite{morozova, petz1, petz2} that unlike the classical case, there exists an infinite family of contractive Riemannian metrics on the set of density operators. This non-uniqueness of the metric can be exploited to find the tightest bound (cf. \cite{geoPRX} for the recent review), but it is an extremely difficult task. The well known candidates for the metric are Fisher (or Bures) metric and Wigner-Yanase metric which is related to Wigner-Yanase skew information \cite{KAROL,geoPRX}.

In this paper, by following the Ref.~\cite{funoQSL}, we  define the distance $D(\rho,\sigma) = \| \rho - \sigma\|_{\rm tr}$, where  $\| X \|_{\rm tr}=\frac{1}{2} {\rm Tr}[\sqrt{X^{\dagger}X}]$ denotes the trace norm. From triangle inequality, it is straightforward  to obtain the following inequality,
\begin{equation}\label{D2}
  \| \rho(\tau) - \rho(0)\|_{\rm tr} \leq \int_0^\tau \| \dot{\rho}(t) \|_{\rm tr} dt,
\end{equation}
which looks here similar to that of Eq. (\ref{D1}), and thus analogically, the speed $v_t$ is nothing but $\parallel\dot{\rho}(t)\parallel_{\rm tr}$.
Note that, trace norm does not correspond to a Riemannian metric but it is contractive, i.e. $ \| \mathcal{E}(\rho) - \mathcal{E}(\sigma)\|_{\rm tr} \leq  \| \rho - \sigma\|_{\rm tr}$, where $\mathcal{E}$ stands for an arbitrary quantum channel. It should be stressed that the trace norm is the most natural measure of distinguishability of quantum states. Moreover, any Markovian evolution satisfies,
\begin{equation}\label{}
  \frac{d}{dt} \| \rho(t) - \sigma(t) \|_{\rm tr} \leq 0 , 
\end{equation}
for any two trajectories starting from $\rho(0)$ and $\sigma(0)$. It shows, that the use of the trace norm to find the bound for the quantum speed limit is fully justified.
Suppose now that $\rho(t)$ satisfies time-local master equation,
\begin{equation}\label{ME}
  \dot{\rho}(t) = \mathcal{L}_t[\rho(t)] =  -\frac{i}{\hbar} [H(t),\rho(t)] + \mathcal{D}_t[\rho(t)],
\end{equation}
with time-local generator $\mathcal{L}_t$ consisting of the Hamiltonian part  and dissipative part governed by $\mathcal{D}_t$.  Following Ref.~\cite{funoQSL} one arrives at the following bound

\begin{equation}
\label{QSL-geo}
\tau\geq\tau_{\rm q1}:= \frac{\parallel\rho(\tau)-\rho(0)\parallel_{\rm tr}}{\langle\parallel \mathcal{L}_t[\rho(t)]\parallel_{\rm tr}\rangle_\tau} .
\end{equation}
After expressing the QSL in this form, authors in the Ref. \cite{funoQSL} further lower bounded this in terms of thermodynamic quantities. Let us consider that
\begin{equation}   \label{sp}
\rho(t)=\sum_i p_{\alpha}(t) \ket{\alpha(t)}\bra{\alpha(t)}
\end{equation}
be an instantaneous spectral decomposition of $\rho(t)$. Define the diagonal part of the dissipator
\begin{equation}\label{diag}
  \mathcal{D}^{\rm diag}_t[\rho(t)]  =  \sum_{\alpha} { |\alpha(t)\>\< \alpha(t)| \mathcal{D}_t[\rho(t)] |\alpha(t)\>\<\alpha(t)|} ,
\end{equation}
and the non-diagonal complement $ \mathcal{D}^{\rm ndiag}_t = \mathcal{D}_t - \mathcal{D}^{\rm diag}_t$, that is,
\begin{equation}\label{ndiag}
  \mathcal{D}^{\rm ndiag}_t[\rho(t)]  =  \sum_{\alpha\neq \beta} { |\alpha(t)\>\< \alpha(t)| \mathcal{D}_t[\rho(t)] |\beta(t)\>\<\beta(t)|},
\end{equation}
with respect to the eigenbasis $|\alpha(t)\rangle$. One finds that
\begin{equation}\label{diag}
  \mathcal{D}^{\rm diag}_t[\rho(t)]  = \sum_\alpha \dot{p}_\alpha (t) |\alpha(t)\>\<\alpha(t)|,
\end{equation}
and
\begin{equation}\label{}
  \mathcal{D}^{\rm ndiag}_t[\rho(t)] = -i/\hbar [H_\mathcal{D}(t),\rho(t)],
\end{equation}
where
\begin{equation}\label{}
  H_\mathcal{D}(t)  = i \hbar \sum_{\alpha\neq \beta} \frac{ |\alpha(t)\>\< \alpha(t)| \mathcal{D}_t[\rho(t)] |\beta(t)\>\<\beta(t)|}{p_\beta(t) - p_\alpha(t)} .
\end{equation}
Finally, one defines
\begin{equation}\label{}
  (\Delta E)^2 = {\rm Tr}[H^2(t) \rho(t)] - ({\rm Tr}[H(t) \rho(t)])^2,
\end{equation}
together with
\begin{eqnarray}\label{}
  (\Delta E_\mathcal{D})^2 &=& {\rm Tr}[H_\mathcal{D}^2(t) \rho(t)] - ({\rm Tr}[H_\mathcal{D}(t) \rho(t)])^2 \nonumber \\ & =&  {\rm Tr}[H_\mathcal{D}^2(t) \rho(t)] .
\end{eqnarray}
If  the time local master equation given by Eq. (\ref{ME}) is a Lindblad type master equation \cite{funoQSL, breuer02}, then the following bound holds \cite{funoQSL}:
\begin{equation}
\label{funo-marko}
\tau \geq  \tau'_{\rm q2} := \frac{\parallel \rho(\tau)-\rho(0)\parallel_{\rm tr}}{\hbar^{-1} \langle \Delta E\rangle_\tau + \hbar^{-1} \langle \Delta E_D\rangle_\tau+\sqrt{\frac{1}{2}\langle\dot{\sigma}\rangle_\tau\langle A\rangle_\tau}},
\end{equation}
where the entropy production rate $\dot{\sigma}$ \cite{EP-1,Ep-2} (and references within) is defined as follows,
\begin{equation}\label{}
  \dot{\sigma} = \dot{S}(t) - \beta \dot{Q}(t) ,
\end{equation}
where $S(t) = - {\rm Tr}[\rho(t)\log \rho(t)]$ stands for the von Neumann entropy, and $\dot{Q}(t) = {\rm Tr}[\dot{\rho}(t) {H}(t)]$. Moreover,
\begin{equation}\label{da}
  A = \sum_{\alpha \neq \beta} {\rm Tr}[ \mathcal{D}_t( |\alpha(t)\>\<\alpha(t)|) |\beta(t)\>\< \beta(t)| \rho(t) ] ,
\end{equation}
denotes the quantum analog of so called dynamical activity \cite{dynamical-1,dynamical-2,dynamical-3,dynamical-4,PRL-shiraishi}.  A more comprehensive expression will be provided later in discussion. Our goal in this paper is to obtain a similar  bound for a general time-local master equation and thus extending it for non-Markovian dynamics too.

\section{Non-Markovian dynamics and renormalized entropy}
\label{sec-2}
Any time local master equation can be cast in the canonical form \cite{hall},
\begin{equation}
\mathcal{L}_t[\rho(t)] = -i/\hbar [H(t),\rho(t)] + \mathcal{D}_t[\rho(t)],
\end{equation}
with a unitary part (the commutator) and a dissipative part,
\begin{eqnarray}
 \mathcal{D}_t[\rho(t)]  = \sum_i\gamma_i(t)\Big[A_i(t)\rho A_i^{\dagger}(t)-\frac{1}{2}\{A_i^{\dagger}(t)A_i(t),\rho(t)\}\Big]. \nonumber\\
\end{eqnarray}
Here the noise operators $A_i(t)$ satisfy ${\rm Tr}[A_i(t)]=0$ and ${\rm Tr}[A_i^{\dagger}(t)A_j
(t)]=\delta_{ij}$. The formal solution to Eq.~(\ref{ME}) reads as  $\rho(t) = \Lambda_t \rho(0)$, where the dynamical map $\Lambda_t$ has the following form
\begin{equation}\label{}
  \Lambda_t = \mathcal{T} \exp\left( \int_0^t \mathcal{L}_u du \right).
\end{equation}
Here `$\mathcal{T}$' stands for the time-ordering operator. One calls quantum evolution Markovian if the corresponding dynamical map $\Lambda_t$ is CP-divisible, that is, for any $t > s$ the following divisibility property holds $\Lambda_t = V_{t,s} \Lambda_s$ and the propagator $V_{t,s}$ is completely positive and trace-preserving \cite{RHP}. There are other concepts of non-Markovian evolution (cf. recent reviews \cite{RHP,rivas14quantum,b-review,NM3,NM4}). In particular, often non-Markovianity is related to information backflow \cite{BLP}, that is,  there exists a pair of initial states  $\{\rho_1,\rho_2\}$ such that $\frac{d}{dt} \| \Lambda_t \rho_1 - \Lambda_t \rho_2\|_1 > 0$ for some $t > 0$. This concept is closely related to so called P-divisibility which is equivalent to the requirement that the propagator $V_{t,s}$ is positive and trace-preserving \cite{Sabrina,PRL-Angel}.
In this regard, it is worth mentioning that Information backflow and CP-divisibility are non-equivalent notions on Markovianity, and, all dynamics which are CP-divisible are also Markovian from Information Backflow approach \cite{rivas14quantum,b-review}.

Now, using the time-local master equation (\ref{ME}) and instantaneous spectral decomposition of $\rho(t)$ from Eq.~(\ref{sp}), it is easy to obtain a Pauli type master equation \cite{leggio} for $p_\alpha(t)$
\begin{equation}
\label{pauli-master}
\dot{p}_\alpha(t)=\sum_\beta[R_{\alpha\beta}(t)p_\beta(t)-R_{\beta\alpha}(t)p_\alpha(t)],
\end{equation}
where the instantaneous transition rate $R_{\alpha\beta}(t)$ between two states $|\alpha(t)\rangle$ and $|\beta(t)\rangle$  is defined as
\begin{equation}\label{}
 R_{\alpha\beta}(t)=\sum_i\gamma_i(t){|\bra{\alpha(t)}A_i(t)\ket{\beta(t)}|}^2 .
\end{equation}
This master equation will be main tool for the rest of our analysis in this paper.
If all the transition rates are positive, the Pauli master equation is nothing but the master equation for a stochastic Markov process. Therefore, Eq. (\ref{pauli-master}) naturally implies a quantum trajectory analogous to the trajectory in classical stochastic process. Specifically, one can consider the set of eigenbasis $\{\ket{\alpha(t)}\}$ as the set of states with probability $p_\alpha(t)$, and the system jumps from one state $\ket{\alpha(t)}$ to another state $\ket{\beta(t)}$ with the transition rate $R_{\alpha\beta}(t)$ at time $t$.

Note that, positivity of the transition rates $R_{\alpha\beta}(t)$ is guaranteed whenever the evolution is P-divisible \cite{bruer-P}. Now, if $\Lambda_t$ is non-Markovian, that is,  at least one of the rate $\gamma_i(t)$ is negative, the transition rates $R_{\alpha\beta}(t)$ might be negative as well. Clearly, there are non-Markovian dynamical maps $\Lambda_t$ for which the rates $R_{\alpha\beta}(t)$ are always non-negative.
Our analysis is aimed to include those scenarios where transition rates can also be temporarily negative.

To describe the entropy production rate in this situation, we follow the framework established in the Ref. \cite{leggio}. Using the Pauli master equation Eq. (\ref{pauli-master}), it can easily be shown that the von-Neumann entropy $S(t)$ satisfies
\begin{equation}
\dot{S}(t)=-\sum_{\alpha\neq\beta}p_\alpha(t) R_{\beta\alpha}(t)\log{\frac{p_\beta (t)}{p_\alpha(t)}}.
\end{equation}
Following the convention of entropy production rate in a stochastic jump process \cite{stochastic-1,stochastic-2}, the above formula  can be split in two terms as,
\begin{equation}
\dot{S}(t)=-\sum_{\alpha\neq\beta}p_\alpha R_{\beta\alpha}\log{\frac{R_{\beta\alpha}}{R_{\alpha\beta}}}+\sum_{\alpha\neq\beta}p_\alpha R_{\beta\alpha}\log{\frac{p_\alpha R_{\beta\alpha}}{p_\beta R_{\alpha\beta}}}.
\end{equation}
From now on, we will be omitting the time dependence for the simplicity of notation. The first term in the above splitting is attributed to the external entropy production rate or entropy flow denoted as $\dot{S}_e(t)$ and the second term, denoted as $\dot{S}_{\rm tot}(t)$ is the total entropy production rate. This is a well established notion \cite{stochastic-1,stochastic-2,enpr1,enpr2,enpr3} of a general markov jump process described by the Pauli master equation. Now, the above formulation is purely mathematical and can be applied to any system whose dynamics is governed by the above master equation. Physical meaning behind such identifications come when one provides additional detailing of the framework, i.e., the Hamiltonian of the system, heat baths with well defined thermodynamic variables (like temperature, chemical potential) etc.  For example, consider a system interacting with a heat bath, obeying standard
Lindblad dynamics \cite{funoQSL}. If the stationary bath state is a thermal state ($e^{-\beta H_B}/{\rm Tr}[e^{-\beta H_B}]$), where $H_B$ is the bath Hamiltonian) with inverse temperature $\beta$, bath correlation functions obey the Kubo-Martin-Schwinger (KMS) condition, which is also known as detailed balance
condition. Owing to this condition, the external entropy flow can be identified as rate of the
heat exchange between the system and the bath and as a consequence $\dot{S}_{\rm tot}(t)$ is identified as the
total or irreversible entropy production rate.

Now, as pointed out in the Ref. \cite{leggio}, having some of the transition rates  negative, we face two problems. First, logarithm of a negative number is ill-defined and second, with Pauli master equation involving negative transition rates, it is not possible to give a pure state single trajectory description. To deal with the first problem, authors in the Ref. \cite{leggio} wrote down the splitting of $\dot{S}(t)$ in the following form,
\begin{equation}
\dot{S}(t)=\dot{\mathcal{S}}_e(t)+\dot{\mathcal{S}}_{\rm tot}(t),
\end{equation}
where the redefined quantities have the expression according to  the Ref. \cite{leggio} as follows,
\begin{align}
&\dot{\mathcal{S}}_e(t)=-\sum_{\alpha\neq\beta}p_\alpha R_{\beta\alpha}\log\frac{|R_{\beta\alpha}|}{|R_{\alpha\beta}|},\\
&\dot{\mathcal{S}}_{\rm tot}(t)=\sum_{\alpha\neq\beta}p_\alpha R_{\beta\alpha}\log{\frac{p_\alpha|R_{\beta\alpha}|}{p_\beta|R_{\alpha\beta}|}}.\nonumber\\
\end{align}
This splitting of $\dot{S}(t)$ in the above form holds as only $\log\frac{|R_{\beta\alpha}|}{|R_{\alpha\beta}|}$ has been added and subtracted in the sum.
Regarding the second problem, one has to redefine the transition rates in such a way that they are always positive without changing the  Pauli master equation. The renormalized transition rates are defined as \cite{leggio}
\begin{equation}
T_{\alpha\beta}=R_{\alpha\beta}^{\rm M}+\frac{p_{\alpha}}{p_\beta}R_{\beta\alpha}^{\rm NM},
\end{equation}
with $R_{\alpha\beta}^{\rm M}=\frac{1}{2}[|R_{\alpha\beta}|+R_{\alpha\beta}]$ and $R_{\alpha\beta}^{\rm NM}=\frac{1}{2}[|R_{\alpha\beta}|-R_{\alpha\beta}]$ denoted as the Markovian and non-Markovian contributions respectively.
Renormalized transition rates $T_{\alpha\beta}$ boils down to the original transition rates $R_{\alpha\beta}$ when all $R_{\alpha\beta}$'s are positive. With these renormalized positive transition rates, the Pauli master equation Eq. (\ref{pauli-master}) can be re-written as,
\begin{equation}
\label{effective-Pauli}
\dot{p}_\alpha(t)=\sum_\beta[T_{\alpha\beta}p_\beta(t)-T_{\beta\alpha}p_\alpha(t)],
\end{equation}
which is equivalent to the Eq. (\ref{pauli-master}), if the initial condition $p_\alpha(0)$ for Eq. (\ref{effective-Pauli}) is same as for Eq. (\ref{pauli-master}). In writing this way, the advantage is that this master equation now describes both Markovian and non-Markovian dynamics although the effective transition rates are positive. But as $T_{\alpha\beta}$ depends upon $p_\alpha$'s, in the non-Markovian regime different trajectories are no longer independent \cite{leggio}.
One can now define the effective Markovian like total entropy production and entropy flow rate \cite{leggio}
\begin{align}
&\dot{S}_{\rm tot}^{\rm M}=\sum_{\alpha\neq\beta}p_\alpha T_{\beta\alpha}\log{\frac{p_\alpha T_{\beta\alpha}}{p_\beta T_{\alpha\beta}}},\\
& \dot{S}_{\rm e}^{\rm M}=-\sum_{\alpha\neq\beta}p_\alpha T_{\beta\alpha}\log{\frac{T_{\beta\alpha}}{T_{\alpha\beta}}}.
\end{align}
This identification now separates the term which is irreducibly non-Markovian as
\begin{equation}
\dot{S}_{\rm tot}^{\rm NM}=\dot{S}_{\rm tot}^{\rm M}-\dot{\mathcal{S}}_{\rm tot}=\dot{\mathcal{S}}_{\rm e}-\dot{S}_{\rm e}^{\rm M}=\sum_{\alpha\neq\beta}p_\alpha T_{\beta\alpha}\log{\frac{T_{\beta\alpha} |R_{\alpha\beta}|}{|R_{\beta\alpha}| T_{\alpha\beta}}}.
\end{equation}
$\dot{S}_{\rm tot}^{\rm NM}$ is clearly zero if all renormalized transition rates are same as the original ones. This happens, when we don't need to renormalize, i.e., when all the transition rates are positive.
As discussed earlier, this happens if and only if the dynamics is P-divisible. $\dot{\mathcal{S}}^{\rm M}_{\rm tot}$ is only a part of total entropy production rate. With additional physical input, the thermodynamic framework can be justified for the total entropy production rate and hence $\dot{\mathcal{S}}^{\rm M}_{\rm tot}$ can be regarded as the resultant of an additional term $\dot{S}_{\rm tot}^{\rm NM}$ arising due to the correlation among different trajectories as they are not independent in the non-Markovian scenario due to the memory effects. This can be thought as new face of second law in the same line of the Ref. \cite{enpr3}, where the total entropy production rate has been splitted in three different forms. Moreover, as shown in the Ref. \cite{leggio}, its single trajectory counterpart gives a fluctuation relation for non-Markovian dynamics.

This formalism will now be applied to evaluate the speed limit in terms of thermodynamic quantities for the most general time-local master equation.

\section{QSL  IN TERMS OF THERMODYNAMIC QUANTITIES}
\label{sec-3}
Exploiting  Eq.~(\ref{QSL-geo}), we now evaluate the QSL for the time-local master equation~(\ref{ME}) with the formalism set in the previous section. This reduces to the fact that we have to calculate the lower bound of the denominator of Eq.~(\ref{QSL-geo}), which is $\langle\parallel \mathcal{L}_t[\rho(t)]\parallel_{\rm tr}\rangle_\tau$. To evaluate this, we consider the following inequality in Ref.~\cite{funoQSL}
\begin{widetext}
\begin{equation}   \label{3-terms}
\parallel \mathcal{L}_t[\rho(t)]\parallel_{\rm tr}\leq  \frac{1}{\hbar} \parallel [H(t),\rho(t)]\parallel_{\rm tr} + \parallel \mathcal{D}^{\rm diag}_t[\rho(t)]\parallel_{\rm tr} +  \parallel \mathcal{D}^{\rm ndiag}_t[\rho(t)]\parallel_{\rm tr} ,
\end{equation}
\end{widetext}
where $\mathcal{D}^{\rm diag}_t$ and $\mathcal{D}^{\rm ndiag}_t$ are defined in (\ref{diag}) and (\ref{ndiag}), respectively. Now, the essential difference is the diagonal part $\mathcal{D}^{\rm diag}_t[\rho(t)] = \sum_\alpha \dot{p}_\alpha |\alpha\>\<\alpha|$, since $p_\alpha(t)$ satisfies the non-Markovian Pauli rate equation. To calculate the bound on $\parallel \mathcal{D}^{\rm diag}_t[\rho(t)]\parallel_{\rm tr}$ we follow the approach of the Ref. \cite{funoQSL}:

\begin{widetext}
\begin{eqnarray}
\nonumber
\parallel \mathcal{D}_t^{\rm diag}[\rho(t)]\parallel_{\rm tr} = \frac{1}{2}\sum_{\alpha}|\dot{p}_\alpha| &=&\frac{1}{2}\sum_\alpha \abs[\Big]{\sum_\beta [T_{\alpha\beta}p_\beta(t)-T_{\beta\alpha}p_\alpha(t)]}\\
\nonumber
& \leq & \frac{1}{2}\sum_\alpha \sqrt{\sum_\beta\frac{{[T_{\alpha\beta}p_\beta(t)-T_{\beta\alpha}p_\alpha(t)]}^2}{[T_{\alpha\beta}p_\beta(t)+T_{\beta\alpha}p_\alpha(t)]}\Big(\sum_\gamma [T_{\alpha\gamma}p_\gamma(t)+T_{\gamma\alpha}p_\alpha(t)]\Big)}\\
& \leq & \frac{1}{2}\sqrt{\sum_{\alpha\neq\beta}\frac{{[T_{\alpha\beta}p_\beta(t)-T_{\beta\alpha}p_\alpha(t)]}^2}{[T_{\alpha\beta}p_\beta(t)+T_{\beta\alpha}p_\alpha(t)]}\Big(\sum_{\alpha\neq\gamma}[T_{\alpha\gamma}p_\gamma(t)+T_{\gamma\alpha}p_\alpha(t)]\Big)}.
\end{eqnarray}

We have used Cauchy Schwartz inequality to obtain the second and third line of the above equation.
Finally, using the inequality $2{(a-b)}^2/(a+b)\leq (a-b)\log{(a/b)}$ \cite{funoQSL} for non-negative $a$ and $b$, one gets
\begin{eqnarray}
\nonumber
\parallel \mathcal{D}_t^{\rm diag}[\rho(t)]\parallel_{\rm tr} &\leq &  \frac{1}{2}\sqrt{\frac{1}{2}\sum_{\alpha\neq\beta}[T_{\alpha\beta}p_\beta(t)-T_{\beta\alpha}p_\alpha(t)]\log{\frac{T_{\alpha\beta}p_\beta(t)}{T_{\beta\alpha}p_\alpha(t)}}\Big(\sum_{\alpha\neq\gamma} [T_{\alpha\gamma}p_\gamma(t)+T_{\gamma\alpha}p_\alpha(t)]\Big)}  \leq \sqrt{\frac{\dot{S}_{\rm tot}^{\rm M}A}{2}},
\end{eqnarray}
\end{widetext}
{where},
\begin{eqnarray}
\nonumber
 \dot{S}_{\rm tot}^{\rm M} &=& \sum_{\alpha\neq\beta}p_\beta(t) T_{\alpha\beta}\log{\frac{p_\beta(t) T_{\alpha\beta}}{p_\alpha(t) T_{\beta\alpha}}} \\
 &=&\frac{1}{2}\sum_{\alpha\neq\beta} [T_{\alpha\beta}p_\beta(t)-T_{\beta\alpha}p_\alpha(t)]\log{\frac{T_{\alpha\beta}p_\beta(t)}{T_{\beta\alpha}p_\alpha(t)}} \nonumber\\
\end{eqnarray}
and,
\begin{eqnarray}
A=\frac{1}{2}\sum_{\alpha\neq\gamma} [T_{\alpha\gamma}p_\gamma(t)+T_{\gamma\alpha}p_\alpha(t)].
\end{eqnarray}
The quantity  `$A$' is identified as the quantum dynamical activity as mentioned earlier in Eq. (\ref{da}). From the expression it is evident that $A$ quantifies the frequency of jump between $p_{\alpha}(t)$'s.
Two remaining terms in (\ref{3-terms}) are evaluated in the same way as in \cite{funoQSL}. As trace norm is non increasing under a CPTP (completely positive and trace preserving) map, by introducing $\ket{\rho}$ as the the purification of $\rho(t)$ and the partial trace operation (a CPTP map) $\Phi$, such that $\Phi(\bar{H}(t))=H(t)$, where $\bar{H}(t)=\mathds{1}\otimes H(t)$ is the natural extension of $H(t)$ to a higher dimensional space, we can write,
\begin{equation}
\frac{1}{\hbar}\parallel [H(t),\rho(t)]\parallel_{\rm tr} \leq \frac{1}{\hbar}\parallel[\bar{H}(t),\ket{\rho}\bra{\rho}]\parallel_{\rm tr}=\frac{1}{\hbar}\Delta E,
\end{equation}
where, ${(\Delta E)}^2=\bra{\rho}{\bar{H}}^2\ket{\rho}-{\bra{\rho}\bar{H}\ket{\rho}}^2={\rm Tr}[\rho(t)H(t)^2]-{({\rm Tr}[\rho(t)H(t)])}^2$  is the energy fluctuation. Similarly, for the non-diagonal part of the dissipator,
\begin{equation}
\parallel \mathcal{D}_t^{\rm ndiag}[\rho(t)]\parallel_{\rm tr}= \frac{1}{\hbar} \parallel [H_\mathcal{D},\rho(t)]\parallel_{\rm tr}\leq \frac{1}{\hbar}\Delta E_\mathcal{D},
\end{equation}
where ${(\Delta E_\mathcal{D})}^2={\rm Tr}[\rho(t)H_\mathcal{D}^2]-{({\rm Tr}[\rho(t)H_\mathcal{D}])}^2$.  As $\Delta E$ can be interpreted as the velocity of the isolated system, analogically, $\Delta E_{\mathcal{D}}$ can be interpreted as the velocity due to bath induced unitary dynamics \cite{funoQSL}. \\\\
Now, combining  all the terms, we finally get,
\begin{equation}
\parallel \mathcal{L}_t[\rho(t)]\parallel_{\rm tr}\leq \frac{1}{\hbar} \Delta E + \frac{1}{\hbar}\Delta E_\mathcal{D} + \sqrt{\frac{\dot{S}_{\rm tot}^{\rm M}A}{2}}.
\end{equation}
From now on, we consider $\hbar=1$, unless otherwise stated. Now, by averaging over the time interval $0$ to $\tau$, the speed limit (applying the Cauchy-Schwartz inequality once again) according to the Eq. (\ref{QSL-geo}) can be expressed as,
\begin{equation}
\label{QSL-thermo}
\tau\geq \tau_{\rm q1}\geq\tau_{\rm q2}:=\frac{\parallel \rho(\tau)-\rho(0)\parallel_{\rm tr}}{\langle\Delta E\rangle_\tau + \langle\Delta E_{\mathcal{D}}\rangle_\tau+\sqrt{\frac{1}{2}\langle\dot{S}_{\rm tot}^{\rm M}\rangle_\tau\langle A\rangle_\tau}}.
\end{equation}
Clearly the form of this QSL  bound is similar with the expression in Eq.~(\ref{funo-marko})~\cite{funoQSL}, but it exploits the framework for non-Markovian dynamics. We have been able to find the bound in terms of thermodynamic quantities for any time-local master equation whether it is   Markovian or non-Markovian. We have denoted the new bound as $\tau_{\rm q2}$ to distinguish it from its Markovian counterpart $\tau'_{\rm q2}$ in Eq. (\ref{funo-marko}).

In the subsequent sections, we illustrate this approach with paradigmatic examples of non-Markovian qubit evolution.

\section{Examples: QSL for qubit evolution}
\label{sec-4}
\subsection{Amplitude damping evolution}

In this section, we analyze amplitude damping dynamics \cite{jc2, jc1, breuer02} governed by the following time-local master equation
\begin{equation}
\label{damp_master}
\frac{d\rho(t)}{dt}=\gamma(t)\Big[\sigma_-\rho(t)\sigma_+-\frac{1}{2}\{\sigma_+\sigma_-,\rho(t)\}\Big],
\end{equation}
where,
\begin{equation}
\gamma(t)=\frac{2\gamma_0\lambda\sinh(dt/2)}{d\cosh(dt/2)+\lambda\sinh(dt/2)},
\end{equation}
and $\sigma_{\pm}=\frac{1}{2}(\sigma_x\pm i\sigma_y)$. Here $\gamma_0$ is the coupling strength between the qubit and the environment with a Lorentzian spectral density having spectral width $\lambda$ and $d=\sqrt{\lambda^2-2\gamma_0\lambda}$. In the interaction picture, the solution of this master equation is given by
\begin{equation}
\rho(t)=
\begin{pmatrix}
\rho_{00}{|b_t|}^2 & \rho_{01}b_t\\
\rho_{10}{b^*_t}  & 1-\rho_{00}{|b_t|}^2
\end{pmatrix},
\end{equation}
with $b_t=e^{-\lambda t/2}\Big(\cosh(\frac{dt}{2})+\frac{\lambda}{d}\sinh(\frac{dt}{2})\Big)$. The dynamics is Markovian in the weak coupling regime, specifically for $\gamma_0<\lambda/2$. For this master equation in Eq.~(\ref{damp_master}), the renormalized transition rates for the Pauli master equation read as follows
\begin{equation}
T_{\alpha\beta}=R_{\alpha\beta}^{\rm M}+\frac{p_{\alpha}}{p_\beta}R_{\beta\alpha}^{\rm NM},
\end{equation}
with $R_{\alpha\beta}(t) = \gamma(t){|\bra{\beta(t)}\sigma_-\ket{\alpha(t)}|}^2$, and $\rho(t)=\sum_\alpha p_\alpha (t) \ket{\alpha(t)}\bra{\alpha(t)}$. Since there is no unitary part in the master equation, the QSL in this case simplifies to
\begin{equation}
\label{QSL-example}
\tau_{\rm q2}=\frac{\parallel \rho(\tau)-\rho(0)\parallel_{\rm tr}}{\langle\Delta E_\mathcal{D} \rangle_\tau+\sqrt{\frac{1}{2}\langle\dot{S}_{\rm tot}^{\rm M}\rangle_\tau\langle A\rangle_\tau}}.
\end{equation}
\begin{table}[H]
  \begin{center}
 \begin{tabular}{||c|| c c c c||}
 \hline
 State & $\rho_{00}$ & $\rho_{01}$ & Purity & $l_1$ norm \\ [0.5ex]
 \hline\hline
 1 & 0.3 & 0.45 & 0.985 & 0.90 \\
 \hline
2 &  0.3 & 0.005 & 0.580 & 0.01 \\
 \hline
3 & 0.5 & 0.45 & 0.905 & 0.90 \\
 \hline
4 & 0.5 & 0.005 & 0.500 & 0.01 \\
 \hline
 5 & 1/3 & 0.25+0.3i & 0.861 & 0.781 \\
 \hline
 6 & 0.5 & 0.25 (1+i) & 0.75 & 0.707 \\ [1ex]
 \hline
\end{tabular}
\end{center}
\caption{Initial states are of the form $\rho(0):=
  \begin{pmatrix}
    \rho_{00} & \rho_{01} \\
   \rho_{10}  & \rho_{11}
  \end{pmatrix}
$ in the eigen-basis of $\sigma_z$.}
\label{Table1}
 \end{table} 
\begin{figure}[H]
\begin{center}
\includegraphics[width=0.48\textwidth]{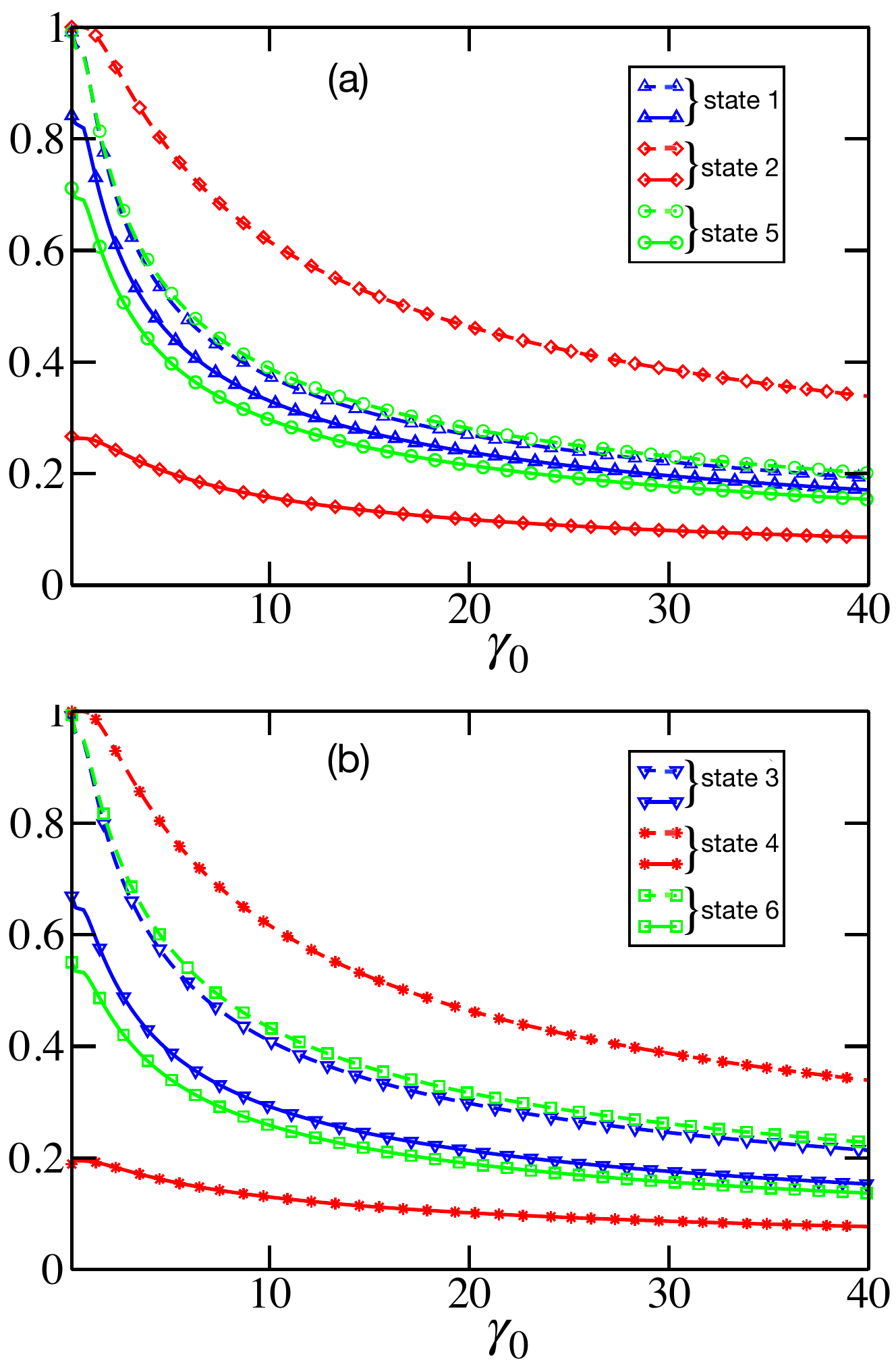}
\caption{(color online) Plot of the ratios $\tau_{q1}/\tau$ and $\tau_{q2}/\tau$ with $\gamma_0$  for the amplitude damping model. Unit of $\gamma_0$ is in $1/{\rm second}$.
The initial states with (a)  different and (b) same diagonal entries are  taken from  Table \ref{Table1}.
 The dashed and solid lines denote the ratios $\tau_{q1}/\tau$ and   $\tau_{q2}/\tau$, respectively. Here $\lambda=1$. Calculations have been performed with logarithm base 2.}
\label{amp-damp}
\end{center}
\end{figure}
To compute the QSL bounds for this model and others, we consider six  different initial states which have been listed
in Table~\ref{Table1}  with their corresponding values of purity~\cite{nielson} and coherence \cite{coh-rev} (quantified in terms of $l_1$ norm). For a quantum state $\rho$, purity is ${\rm Tr}[\rho^2]$ and $l_1$ norm is $\sum_{i\neq j}|\rho_{ij}|$, where $\rho_{ij}=\bra{i}\rho\ket{j}$, w.r.t a particular basis $\{k\}$.
Note that we take into account these states
with contrasting purity and coherences for a better comparison.
One can clearly see that both purity  and coherence of state 1 are large, whereas state 2 has small purity and coherence. Similarly for the states 3 and 4. State 5 and 6 have purity and coherence which are in between the previous states. Plots will show that the bounds $\tau_{\rm q1}$ and $\tau_{\rm q2}$ will depend on the initial states we choose. 
\begin{figure}[h]
\begin{center}
\includegraphics[width=0.49\textwidth]{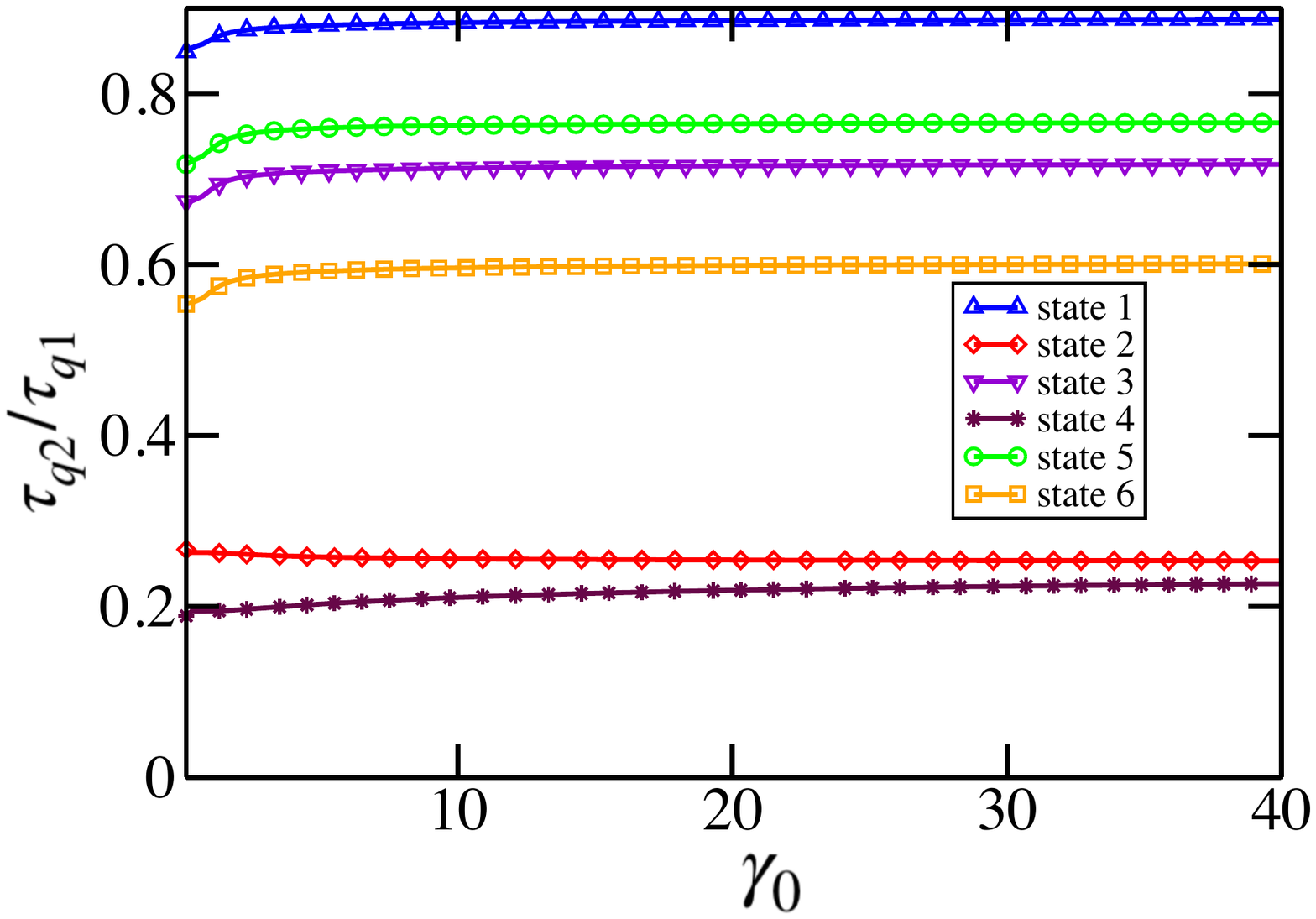}
\caption{(color online) Plot of the ratio $\tau_{\rm q2}/\tau_{\rm q1}$ with coupling strength $\gamma_0$ for amplitude damping model. Unit of $\gamma_0$ is in $1/{\rm second}$. Here $\lambda=1$.
}
\label{ratio-JC}
\end{center}
\end{figure}
In Figures \ref{amp-damp} (a) and \ref{amp-damp} (b), we plot the ratio of the bound $\tau_{\rm q1}$ obtained from the Eq. (\ref{QSL-geo}) and actual time $\tau$ simultaneously with the ratio of the bound $\tau_{\rm q2}$ from the Eq. (\ref{QSL-thermo}) and actual time $\tau$ with the coupling strength $\gamma_0$ for $\lambda=1$. This means that the dynamics is non-Markovian for $\gamma_0>0.5$.
Note that in Fig.~\ref{amp-damp} (a), we consider those initial states for which the diagonal entries are not equal i.e. states 1, 2 and 5, and Fig.~\ref{amp-damp} (b) consists the same diagonal entries states i.e. states 3, 4, and 6.
 Both the plots show that both $\tau_{\rm q1}$ and $\tau_{\rm q2}$ are dependent on the initial states we choose. Roughly, they depend mainly on the coherence  and purity  of the initial density matrices. Larger is the coherence and purity of the state, better is the bound $\tau_{\rm q2}$. For example, both the states 2 and 4 have same coherence, but purity of the state 2 is more than that of the state 4. As a consequence, state 2 gives slightly better bound $\tau_{\rm q2}$ than state 4. Same is true for the states 1 and 3, and also for the states 5 and 6. Though for very low coherence, sensitivity of the bound on purity is less prominent as the bound remains more or less same.
Interestingly, opposite is true for the bound $\tau_{\rm q1}$ obtained from Eq. (\ref{QSL-geo}), which is evident from the above plots. For example, state 2 and state 4 have the lowest coherence and purity, but these two states give best estimate for the bound $\tau_{\rm q1}$, whereas $\tau_{\rm q2}$ gives poor estimate for these states.
Nevertheless, the inequality $\tau_{\rm q1}\geq \tau_{\rm q2}$ always hold as seen in the plots.
For better explanation, we plot the ratio of $\tau_{\rm q2}/\tau_{\rm q1}$ in  Fig.~\ref{ratio-JC}  with the coupling strength $\gamma_0$. This plot clearly shows the conclusion we have made from the previous plots. State 1, which has largest coherence and purity gives the best estimate of $\tau_{\rm q2}$.

\subsection{Pure dephasing model}

In this section, we discuss the pure dephasing channel \cite{dephasing1, dephasing2, dephasing3, dephasing4} describing the time evolution of a qubit interacting with a bosonic reservoir. The microscopic Hamiltonian for this model is
\begin{equation}
H_{\rm tot}=\omega_0\sigma_z+\sum_k\omega_k a_k^{\dagger}a_k+\sum_k\sigma_z(g_ka_k+g_k^*a_k^{\dagger}),
\end{equation}
where $\omega_0$ is the qubit frequency, $\omega_k$ is environment mode frequency, $a_k$ and $a_k^{\dagger}$ are  the annihilation and creation operators respectively,
 and $g_k$ is the coupling constant between the environment mode and the qubit.  This model is exactly solvable.  The master equation for the qubit in the interaction picture can be written as
\begin{equation}
\label{dephasing-eq}
\frac{d\rho(t)}{dt}=\gamma(t)[\sigma_z\rho(t)\sigma_z-\rho(t)].
\end{equation}
With a initial thermal state of the environment, the evolution of the off-diagonal element of the density matrix is given by
\begin{equation}
\rho_{ij}(t)=e^{-\Gamma(t)}\rho_{ij}(0),
\end{equation}
with dephasing factor $\Gamma(t)=2\int_0^t \gamma(t')dt'$. However, the diagonal entries remain same. In the continuum limit, the dephasing rate $\gamma(t)$ is given by
\begin{equation}
\gamma(t)=\int d\omega J(\omega)\coth[\hbar\omega/2k_BT]\sin(\omega t)/\omega,
\end{equation}
where $J(\omega)$ is the reservoir spectral density. Now for Ohmic like spectral densities, a closed form of the dephasing rate can be obtained in the high and low temperature limit with  the form of $J(\omega)$ given below
\begin{equation}
J(\omega)=\frac{\omega^k}{\omega_c^{k-1}}e^{-\omega/\omega_c},
\end{equation}
where $\omega_c$ is the environment cutoff frequency. By varying the parameter $k$,  one can drive from sub-Ohmic environments $(k < 1)$ to Ohmic $(k = 1)$ and super-Ohmic $(k > 1)$ environments, respectively.
Also the  expression for the time-dependent dephasing rate in the zero temperature limit is
\begin{equation}
\gamma(k,t)=\omega_c{[1+{(\omega_ct)}^2]}^{-k/2}\Gamma[k]\sin[k \arctan(\omega_c t)],
\end{equation}
\begin{figure}[t]
\begin{center}
\includegraphics[width=0.48\textwidth]{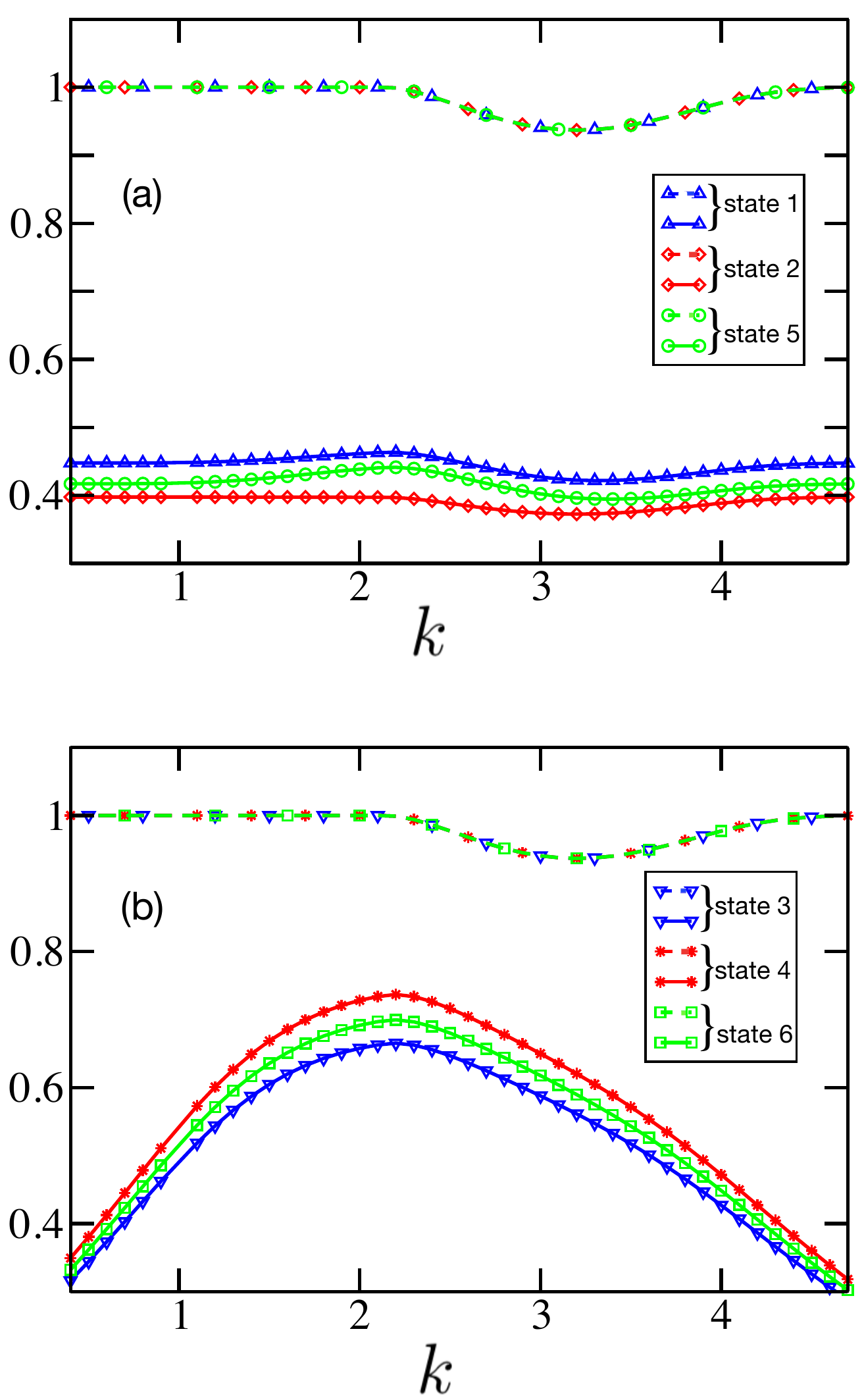}
\caption{(color online) Plot of the ratios $\tau_{q1}/\tau$ and $\tau_{q2}/\tau$ with parameter  $k$ for pure dephasing model. $k$ is dimensionless. The initial states with (a)  different and (b) same diagonal entries are  taken from  Table \ref{Table1}. Here  $\omega_0=0.5$. Dashed and solid lines denote the ratios  $\tau_{q1}/\tau$ and  $\tau_{q2}/\tau$, respectively. Calculations have been performed with logarithm base 2.}
\label{plot-dephasing-1}
\end{center}
\end{figure}
where 􏰘$\Gamma[x]$ is the Euler gamma function.
Note that the dephasing rate can be temporarily negative (indicating the non-Markovian signature) when $k>2$. We employ  the same framework that  we discussed in earlier section to calculate the QSL. For this master equation, the renormalized transition rates for the Pauli master equation read as
\begin{equation}
T_{\alpha\beta}=R_{\alpha\beta}^{\rm M}+\frac{p_{\alpha}}{p_\beta}R_{\beta\alpha}^{\rm NM},
\end{equation}
with $R_{\alpha\beta}(t) =\gamma(t){|\bra{\beta(t)}\sigma_z\ket{\alpha(t)}|}^2$, and $\rho(t)=\sum_\alpha p_\alpha (t) \ket{\alpha(t)}\bra{\alpha(t)}$. As there is no unitary part in the master equation, the QSL can be written as
\begin{equation}
\label{QSL-example}
\tau_{\rm q2}=\frac{\parallel \rho(\tau)-\rho(0)\parallel_{\rm tr}}{\langle \Delta E_\mathcal{D}\rangle_\tau+\sqrt{\frac{1}{2}\langle\dot{S}_{\rm tot}^{\rm M}\rangle_\tau\langle A\rangle_\tau}}.
\end{equation}
Like the previous section, In Fig.~\ref{plot-dephasing-1}, we plot the ratios of $\tau_{\rm q2}/\tau$ and  $\tau_{\rm q1}/\tau$ simultaneously with parameter $k$ for different initial states as given in the Table \ref{Table1}.
We notice an interesting aspect for the pure dephasing case as compared to the amplitude damping model. Unlike the amplitude damping model, the bound $\tau_{\rm q1}$ for pure dephasing (dashed lines in Fig.~\ref{plot-dephasing-1} (a) and Fig. \ref{plot-dephasing-1} (b))  is almost same for all the initial states. But there is a noticeable dependence on the initial state for the bound $\tau_{\rm q2}$. Since the diagonal elements of the density matrices do not change during the time evolution, the nature of the plots in Fig.~\ref{plot-dephasing-1}  for different and equal diagonal elements in the initial density matrices is not same.  Like the previous section, here also one can clearly see that more purity and more coherence yield better $\tau_{\rm q2}$  for the  initial states which have different diagonal elements (Fig.~\ref{plot-dephasing-1} (a)). However, this is just opposite for the cases when the initial states have equal diagonal elements (Fig. \ref{plot-dephasing-1} (b)). Moreover, approximately in the region $1<k<4$, initial states with same diagonal entries give better bound in terms of $\tau_{\rm q2}$.

From these two examples that we discussed above, clearly we can say that both the bounds $\tau_{\rm q1}$ and $\tau_{\rm q2}$ are dependent on the dynamics that we consider, and also the initial states that we choose. For instance,  in the amplitude damping model, if we choose initial state to be a pure state, the entropy production rate diverges, as a consequence the bound $\tau_{\rm q2}$ gets trivial. Therefore the dynamics dictates whether the bound $\tau_{\rm q2}$ in terms of thermodynamic quantities will be a suitable choice or not.

\subsection{Eternally non-Markovian evolution}
In this section, we consider the following time-local qubit generator~\cite{hall}
\begin{equation}\label{eternal-master}
  \mathcal{L}_t(\rho) = \frac 12 \sum_{k=1}^3 \gamma_k(t)( \sigma_k \rho \sigma_k - \rho) ,
\end{equation}
with $\gamma_1=\gamma_2=1$ and $\gamma_3(t) = - \tanh t < 0$. Clearly one of the rates is always negative (for $t>0$), which ensures that the dynamics is non-Markovian, although the corresponding dynamical map found on solving the master equation is completely positive. The authors in Ref.~\cite{hall} call such evolution as {\em eternally non-Markovian}. It can also be seen that the evolution is perfectly P-divisible, which is again a signature of strong non-Markovianity  (it was further generalized in Refs.~\cite{nina} and \cite{kasia} for qudit systems).
\begin{figure}[h]
\begin{center}
\includegraphics[width=0.48\textwidth]{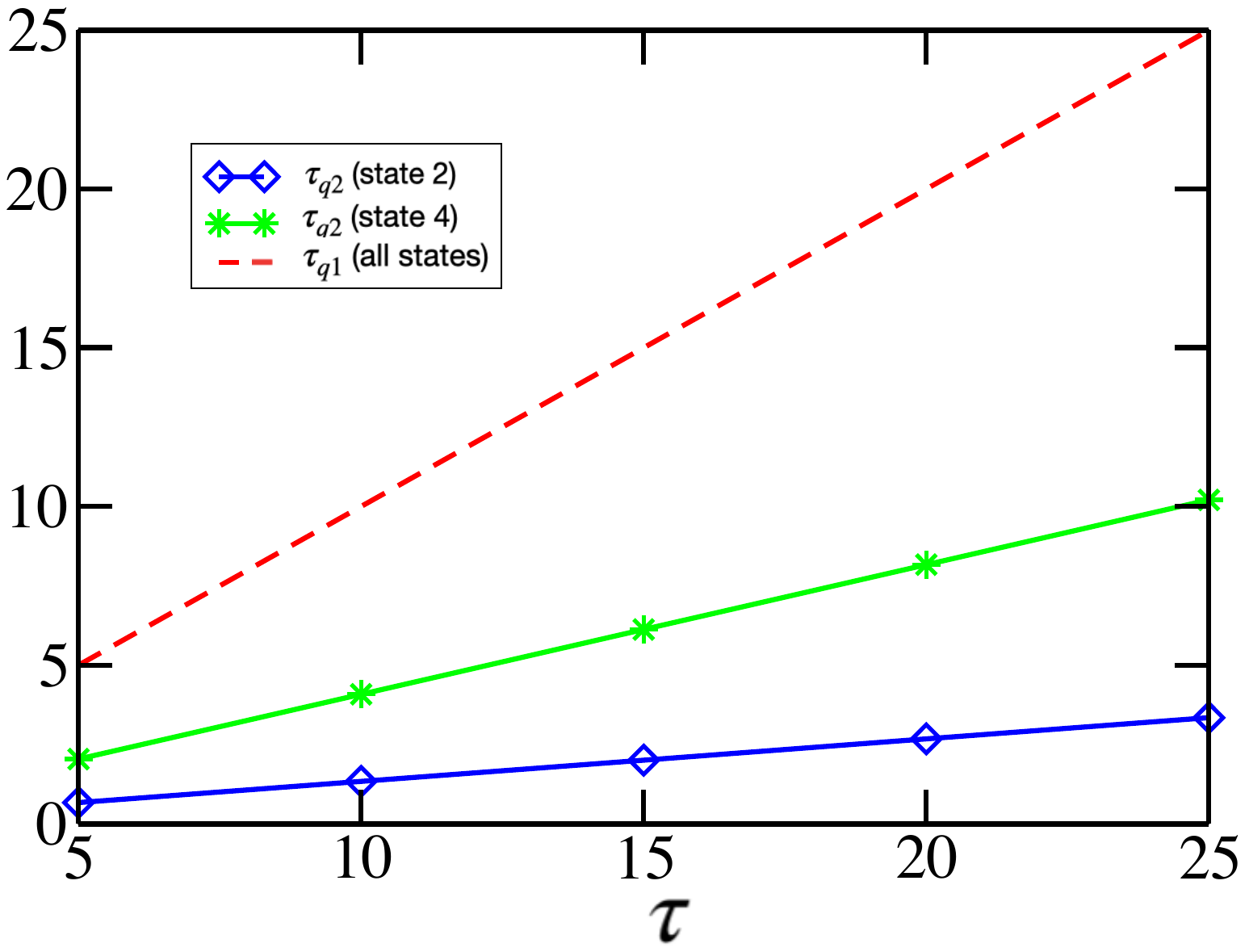}
\caption{(color online) Plot of $\tau_{\rm q2}$ and $\tau_{\rm q1}$ with $\tau$ for eternally non-Markovian evolution. Units of $\tau_{\rm q2}$, $\tau_{\rm q1}$, and $\tau$ are in second.
}
\label{eternal}
\end{center}
\end{figure}
The solution for this master equation \cite{hall,nina,kasia} in terms of the Bloch vector reads as
\begin{equation}
x_j(t)=
  \begin{cases}
    \frac{1}{2}(1+e^{-2t})x_j       & \quad j=1,2 \\
    e^{-2 t}x_j   & \quad j=3
  \end{cases}
\end{equation}
 with $\rho(0)=\frac{1}{2}\big[\openone +x_1\sigma_1+x_2\sigma_2+x_3\sigma_3\big]$. Simple calculation leads to the following formula for the transition rate
\begin{equation}\label{}
  R_{12}(t)  = \left(1 + \frac{x_3^2}{|\mathbf{x}|^2} \right) -  \left(1 - \frac{x_3^2}{|\mathbf{x}|^2} \right) \tanh t,
\end{equation}
where $|\mathbf{x}|^2 = x_1^2 + x_2^2 + x_3^2$. Here $R_{12}(t)  = R_{21}(t)$. Clearly, one has $ R_{12}(t) \geq 0$. Consequently, the renormalized transition rates reduce to these original transition rates. The point is to analyze our result for a strongly non-Markovian dynamics where one has perfectly positive rates.
It is now straightforward to calculate  the bounds $\tau_{\rm q1}$ and $\tau_{\rm q2}$. Interestingly, $\tau_{\rm q1}$ always saturates in this case for all initial states, but as the previous cases, $\tau_{\rm q2}$ depends on the initial states. In Fig. \ref{eternal}, we  plot $\tau_{\rm q1}$ for all initial states and $\tau_{\rm q2}$ for initial states 2 and 4 from Table \ref{Table1} with actual final time $\tau$. We see that the nature of the bound $\tau_{\rm q2}$ is comparable with that of pure dephasing model. For example, like pure dephasing model,  state 2 (blue solid diamond line) gives the lowest and state 4 (green solid star line) gives the best estimate for the bound $\tau_{\rm q2}$. All other states from Table~\ref{Table1}  give (not shown in Fig. \ref{eternal}) the estimates in between these two states.
\section{Conclusion}
\label{sec-7}
In this paper, we have analyzed the QSL for non-markovian dynamics in terms of thermodynamic quantities. Specifically, we have considered the form of the most general time-local master equation. Starting from a geometric definition of quantum speed limit, we have found a lower bound in terms of thermodynamic quantities by exploiting the generator of the dynamics. To accomplish such tasks, we  have rearranged the associated Pauli type master equations in terms of renormalized transition rates \cite{leggio} as the original transition rates can temporarily be negative for non-Markovian dynamics. Renormalized transition rates, in turn, provide an avenue to define effective markovian like entropy production rate and dynamical activity. Next, separately upper bounding the unitary and non-unitary parts of time local generator (its trace-norm specifically) \cite{funoQSL}, we found the QSL $\tau_{\rm q2}$ in terms energy fluctuation, entropy production rate and dynamical activity. This gives a physical intuition to the geometrical terms contained in the expression of QSL for open quantum system not restricted to Markovian dynamics only.

We have illustrated our results numerically considering three physical models, namely, amplitude damping, pure dephasing and eternally non-Markovian models. The first two models are not P-divisible, thus containing temporarily negative transition rates in the associated Pauli type master equation. Our framework is specifically aimed to include these types of models. The last model e.g. eternally non-Markovian model is P-divisible implying all the transition rates are positive. Renormalized transition rates in this case reduces to the original transition rates of the Pauli master equation. Numerical studies showed that the new QSL found in terms of thermodynamic quantities is very much dependent on the dynamics we consider. Not only that for a particular dynamics, its tightness is determined by the initial state too. Purity and coherence of the initial state has a great impact on the tightness of the bound. For example amplitude damping model gives the best estimate of the bound among the three when the initial state has high purity and large coherence. On the other hand, the original geometric bound $\tau_{\rm q1}$ we started with, is almost always saturated for dephasing and eternal non-Markovian model, whereas, for amplitude damping model, it is not and also it depends on the initial state one chooses.
In short, the dynamics and the initial state have a crucial role to play both for the original geometric bound $\tau_{\rm q1}$ and the bound $\tau_{\rm q2}$, which is written in terms of the thermodynamic quantities.
A vast variety of the physical processes being non-Markovian, our results can be relevant to the modern quantum technologies.
\section*{Acknowledgements}


The work was supported by the Polish National Science Centre projects No. 2018/30/A/ST2/00837.

\bibliography{QSL-NM}
\end{document}